\documentclass[aps, prr, amsmath,amssymb, preprint, superscriptaddress 
]{revtex4-2}
\usepackage{graphicx}
\usepackage{epstopdf}
\usepackage{dcolumn}
\usepackage{bm}
 \usepackage{here} 
\usepackage[utf8]{inputenc}
\usepackage[T1]{fontenc}
\usepackage{mathptmx}
\usepackage{caption}
 
\begin{document}
\title{Piezoelectricity in Nominally Centrosymmetric Phases}
\author{Oktay Aktas}
\email[email:]{oktayaktas@xjtu.edu.cn}
\affiliation{State Key Laboratory for Mechanical Behavior of Materials\&Materials Science and Engineering, Xi'an Jiaotong University, Xi'an 710049, China}

\author{Moussa Kangama}
\affiliation{State Key Laboratory for Mechanical Behavior of Materials\&Materials Science and Engineering, Xi'an Jiaotong University, Xi'an 710049, China}
\author{Gan Linyu}
\affiliation{State Key Laboratory for Mechanical Behavior of Materials\&Materials Science and Engineering, Xi'an Jiaotong University, Xi'an 710049, China}
\author{Gustau Catalan}
\affiliation{ICREA and ICN2–Institut Catala de Nanociencia i Nanotecnologia, Campus UAB, 08193 Bellaterra (Barcelona), Catalonia}
\author{Xiangdong Ding}
\email[email:]{dingxd@xjtu.edu.cn}
\affiliation{State Key Laboratory for Mechanical Behavior of Materials\&Materials Science and Engineering, Xi'an Jiaotong University, Xi'an 710049, China}

\author{Alex Zunger }
\affiliation{University of Colorado, Energy Institute, Boulder, Colorado 80309 USA}
\author{Ekhard K. H. Salje}
\email[email:]{ekhard@esc.cam.ac.uk}
\affiliation{Department of Earth Sciences, University of Cambridge, Downing Street, Cambridge CB2 3EQ, UK}
\date{\today}
\begin{abstract}
  Compound phases often display properties that are symmetry-forbidden relative to their nominal, average crystallographic symmetry, even if extrinsic reasons (defects, strain, imperfections) are not apparent. Specifically, breaking the macroscopic inversion-symmetry of a centrosymmetric phase can dominate or significantly change its observed properties while the detailed mechanisms and magnitudes of the deviations of symmetry-breaking are often obscure. Here, we choose piezoelectricity as a tool to investigate macroscopic inversion symmetry breaking in nominally centrosymmetric materials as a prominent example and measure Resonant Piezoelectric Spectroscopy (RPS) and Resonant Ultrasound Spectroscopy (RUS) in 15 compounds, 18 samples, and 21 different phases, including unpoled ferroelectrics, paraelectrics, relaxors, ferroelastics, incipient ferroelectrics, and isotropic materials with low defect concentrations, i.e. NaCl, fused silica, and CaF\textsubscript{2}. We exclude the flexoelectric effect as a source of the observed piezoelectricity yet observe piezoelectricity in all nominally cubic phases of these samples. By scaling the RPS intensities with those of RUS, we calibrate the effective piezoelectric coefficients using single crystal quartz as standard. Using this scaling we determine the effective piezoelectric modulus in nominally non-piezoelectric phases, finding that the ‘symmetry forbidden’ piezoelectric effect ranges from $ \sim $ 1 pm/V to 10$^{-5}$ pm/V ($ \sim 0.5 \%$  to $\sim 2 \times 10^{-7} \%$  of the piezoelectric coefficient of poled ferroelectric lead zirconate titanate). The values for the unpoled ferroelectric phase are only slightly higher than those in the paraelectric phase. The extremely low coefficients are well below the detection limit of conventional piezoelectric measurements and demonstrate RPS as a convenient and ultra-highly sensitive method to measure piezoelectricity. We suggest that symmetry-breaking piezoelectricity in nominally centrosymmetric materials and disordered, unpoled ferroelectrics is ubiquitous.
\end{abstract}
\maketitle

\section{Introduction}
 Along with chiral dichroism, second harmonic generation and the Rashba spin splitting, the classic effects of piezoelectricity and pyroelectricity belong to a group of functionalities enabled by specific crystal class (CC) symmetries. Both effects require an absence of inversion center, i.e. belonging to a non-centrosymmetric (NCS) crystal class. In addition, whereas pyroelectricity is restricted exclusively to polar crystal classes, piezoelectricity is allowed both in polar symmetries as well as in non- polar symmetries, (with the exception of the nonpolar CC of type O that forbids piezoelectricity). Remarkably, piezoelectricity and pyroelectricity were recently observed in nominally centrosymmetric phases such as the cubic paraelectric phases of certain oxides \cite{biancoli2015, hashemi2016, garten2015, wieczorek2006, aktas2013bto, nataf2020, riemer2020, aktas2018, frenkel2017, salje2013sto}. Piezoelectricity was demonstrated in several well-known materials, such as paraelectric phase of BaTiO\textsubscript{3 }and its solid solutions, ferroelastic LaAlO\textsubscript{3} and SrTiO\textsubscript{3}, and relaxor ferroelectric lead magnesium niobate (PMN) \cite{garten2015, wieczorek2006, aktas2013bto, aktas2018, yokota2020, salje2013sto, frenkel2007, frenkel2017} raising much attention both in the scientific and engineering communities \cite{simons2018, frenkel2017, seidel2016, salmani2020, nataf2020}.

Some of the sightings of piezo effects in centrosymmetric phases are perhaps less than compelling, being likely \textit{false positive }determination of the nominal cubic phase. Here,the centrosymmetric phase may be the para phase of a ferroelectric material, where the phase transition from the ferroelectric phase is not properly completed. Such ‘false positive’ observations thus do not necessarily pertain to nominally cubic phases (e.g. when the measurement temperatures are just below the para-to ferro transition point) and thus the appearance of piezoelectricity does not pose a puzzle. Other mechanisms, once thought to explain the observation of piezo effects will be argued below not to be plausible. For example, one needs to be careful to exclude both flexoelectricity and surface piezoelectricity as experimental errors which are common in heavily strained samples but do not lead to piezoelectricity. 

The remaining data and observations of piezoelectricity in nominal cubic materials are broadly divided into two causes:

\textit{(a) Ferroelastic domains within ferroelastic phases and ferroelastic local domains in the paraphrase}: These are related to the ordering of spontaneous strain in ferroelastic materials, which are mechanical analogs of ferroelectrics (spontaneous polarization) and ferromagnets (spontaneous magnetization) \cite{lloveras2012,salje2014,ren2012,viehland2014,bratkovsky1994}. In the ferroelastic phase, twin walls, i.e. boundaries separating ferroelastic domains (or strain states) were shown to be generally polar \cite{nataf2020}. Above the transition temperature, signatures of the ferroelastic phase (also called precursors) occur as strain fluctuations (in the paraelastic phase), leading to structural heterogeneity \cite{lloveras2012, salje2014, ren2012, kartha1991}. These include tweed patterns that are seen in electron diffraction experiments as cross-hatched patterns and were observed to have lengths of 100-2000 angstroms \cite{viehland2014}. Such heterogeneities can be a result of elastic anisotropy, the interaction of local strain with another property, such as octahedral tilting in some perovskite oxides and halides, polarization in incipient ferroelectrics, or compounds whose ferroelectric or ferromagnetic phase is simultaneously ferroelastic \cite{lloveras2012,bratkovsky1994,viehland2014,barone2014}.

\textit{(b)Ferroelectric-like local polar structures within the paraelectric phase:} These give rise to \textit{intrinsic} piezoelectricity although some defect induced polarity may play a role as extrinsic stimulus \cite{bussmannholder2009, lloveras2012, ren2012, roleder2012, bussmannholder2014}. In this context, we also include ferroelectric precursors, which occur in the paraelectric phase and exhibit locally the structural features which define the ferroelectric phase \cite{bussmannholder2009, lloveras2012,ren2012,roleder2012,bussmannholder2014}. Local polar structures have recently been observed by electron microscopy in the paraelectric phase of ferroelectric BaTiO\textsubscript{3} and its solid solution with ferroelastic/incipient ferroelectric SrTiO\textsubscript{3}with sizes of 2-4 nm \cite{benjan2020, tsuda2016}. This class includes locally polar structures in relaxors (or relaxor ferroelectrics), which remain cubic down to absolute zero \cite{cowley2011, fu2012}. In the relaxor literature, they are generally referred to as polar nano regions (PNR’s) \cite{cowley2011, fu2012, riemer2020}. The size of PNR’s ranges from several nm to 20 nm in the case of well-known relaxor lead magnesium niobate (PMN) \cite{fu2012, eremenko2019}.

Length scales of above mentioned local polar structures can extend from some sub-nano meter to, in some cases, nearly a micron.\cite{nataf2020,benjan2020,tsuda2016,viehland2014} We refer to all of these various structures as polar nanostructures.

\textit{Extrinsic vs intrinsic reasonings:} All effects in (a) and (b) could include intrinsic mechanisms (i.e. characteristics of the ideal, pristine bulk effects) as well as extrinsic mechanisms, and it is not always possible to distinguish between them experimentally. Examples of \textit{extrinsic effects} include defect-gradients formed during high temperature synthesis of polycrystals (ceramics) \cite{biancoli2015}. Indeed, in order to observe a macroscopic piezoelectric effect under the local polarity perspective it is necessary that the local piezoelectric tensor components d\textsubscript{ik}(r) do not self-compensate inside the sample, e.g. their sum over all sites has to be non-zero. This can be pictured by an ‘effective bias field’ which prevents self-compensation, which would be expected in the thermodynamic limit and random distributions of the tensor components \cite{biancoli2015,salje2016flexo, schiaffino2017}. Current understanding is that this bias may come from defect-gradients as an extrinsic factor formed during high temperature synthesis of polycrystals (ceramics) \cite{biancoli2015}. The same mechanism has been proposed to occur during growth of single crystals \cite{biancoli2015}. Other mechanisms that suggest an external mechanism have also been reported \cite{hashemi2016, garten2015, bersuker2015}. However, there is also a scenario that will retain a net polarization even if the local polar entities are not subjected to an effective bias field. Such are the simulations \cite{salje2016flexo} that suggest that a collection of polar entities (i.e., polar nanostructures) in materials with no external defects can also lead to a net polarization. In this case, nucleation of the first polar nanostructure biases the rest of the material, rendering it polar and piezoelectric. 

Considering that the mechanisms (a)-(b) raised previously regarding the effective loss of inversion symmetry leading to piezoelectricity in nominally cubic para phases were generic not specific, and that the data are not always uniformly certain and may include real-world unavoidable defects, led us to systematically measure the piezoelectric effect in nominally centrosymmetric materials. Whereas piezoelectricity is a very old effect (discovered in 1880 by Curie brothers) and is indispensable with a vast number of technological applications \cite{acosta2017, mckinstry2018}, finding a new twist to it is unusual and exciting physics. We measured Resonant Piezoelectric Spectroscopy (RPS)\cite{aktas2014kto, salje2013sto,aktas2013bto} of five unpoled ferroelectric compounds (characterized by ferroelectric domains) and their paraelectric phases (with local polar structures including ferroelectric precursors), ferroelastic LaAlO$_3$ (with polar twin walls), four relaxor ferroelectrics (with polar nano regions, PNR’s) above their freezing temperature, incipient ferroelectrics KTaO$_3$ and SrTiO$_3$ with no known polar clusters at room temperature, and materials with low concentrations of defects, namely, NaCl, CaF$_2$, and silica glass. By using piezoelectric quartz single crystal as standard, we first determine the strain generated in Resonant Ultrasound Spectroscopy (RUS) and
calibrate the piezoelectric response detected by Resonant Piezoelectric Spectroscopy (RPS) on poled ferroelectrics lead zirconate titanate (PZT-5H) and LiNbO$_3$, and paraelectric SrTiO$_3$. Then, we show that nominally centrosymmetric phases of materials with local polar entities and chemical defects, like dopants, possess measurable piezoelectric effects that vary by seven orders of magnitude in comparison with poled ferroelectrics that have high piezoelectric coefficients. These values range from those comparable to that of quartz to 10 attometer/volt, which is 3 orders of magnitude below the detection limit of conventional piezoelectric measurements. Figure 1 is discussed in detail in this paper. In light of these results we advocate the idea that piezoelectricity is a common phenomenon in nominally centrosymmetric materials and unpoled ferroelectrics.

\begin{figure}[h!]
			\includegraphics[width=8.6cm]{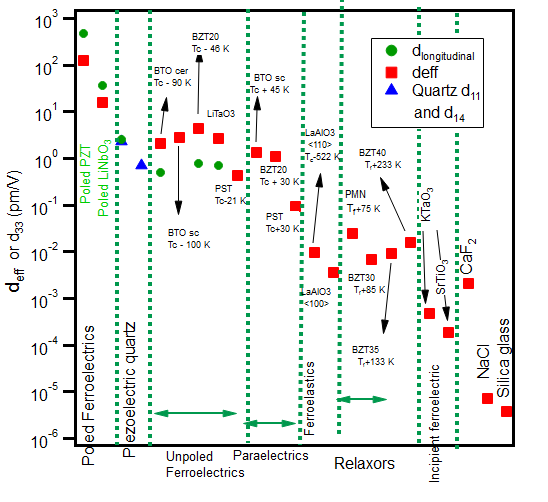}
		\caption{Piezoelectricity in poled ferroelectrics, unpoled ferroelectrics, paraelectric phases of ferroelectrics, nominally centrosymmetric materials, and silica glass. The effective piezoelectric coefficient d\textsubscript{eff}, of each sample, as determined by combined RPS and RUS measurements, corresponds to a coefficient that receives contributions from individual piezoelectric coefficients of the sample. Green circles correspond to longitudinal piezoelectric coefficients measured by a Berlincourt (d\textsubscript{33}) meter. Triangles are piezoelectric coefficients d\textsubscript{11} and d\textsubscript{14} of quartz reported by Ref. \cite{bechmann1958}.}
\end{figure}

\section{METHODS}

Measurements were performed on 13 compounds (15 different samples). Five samples are ferroelectric: a BaTiO\textsubscript{3}single crystal, a BaTiO\textsubscript{3} ceramic, a LiTaO$_3$ single crystal, ceramics of (1-x) BaTiO\textsubscript{3}-x BaZrO$_3$ solid solution with ferroelectric properties (x = 0.20). Relaxors are (1-x) BaTiO\textsubscript{3}-x BaZrO$_3$  with (x = 0.30, 0.35, 0.40) and lead magnesium niobate ceramic (PMN). Nominally non-ferroelectrics are LaAlO\textsubscript{3} single crystals ([001]- and [110]-oriented) in their ferroelastic phase, a single crystal of KTaO$_3$, quartz single crystal (X-cut), NaCl single crystal, CaF$_2$, and silica glass. Ferroelectric, ferroelastic, relaxor, or isotropic behavior of each sample and characteristic temperatures are listed in Table 1. BaTiO\textsubscript{3} and BaTiO$_3$- BaZrO$_3$ ceramics were fabricated by conventional solid-state reaction method with starting chemicals of BaTiO$_3$ (99.9$\%$ ), BaZrO$_3$ (99$\%$). The calcination was performed at 1350 $ ^{\circ}C $  and sintering was done at 1450$ ^{\circ}C$  in air. These samples were preliminarily characterized by x-ray diffraction. A PMN ceramic was provided by Niall Donnelly. The PbSc$_{0.5}$Ta$_{0.5}$O$_3$ (PST) sample is the same as used in an earlier study \cite{ganlinyu2019}. A NaCl single crystal was purchased from Ted Pella Inc. while other crystals and fused silica were purchased from Hefei Kejing Materials Technology. The samples were rectangular parallelepiped platelets with surface areas ranging from 16.98 to 100 mm$^2$ and thicknesses in the range of 0.1 to 1.17 mm. Sample dimensions and orientation of single crystals are listed in the Supplementary material.

Direct piezoelectric constant measurements were done with a quasi-static d$_{33}$ meter (IACAS, ZJ-4AN). For RUS and RPS measurements (Fig. 2), signal generation and detection were achieved via an HF2LI function generator/lock-in amplifier unit (Zurich Instruments). All measurements were performed with PZT transducers with the exception on the BaTiO$_3$ single crystal, for which LiNbO$_3$ transducers were used, and 0.8 (BaTiO$_3$)-0.2 (BaZrO$_3$) in the paraelectric phase for which PIN-PMN-PT transducers were used. Each sample was mounted between two transducers along its corners, which maximize their resonance intensities.  RPS and RUS measurements of each sample were performed without changing the sample position. Measurements in the paraelectric phases of BaTiO$_3$, BZT20, and PST and ferroelectric phases of PST were performed in a nitrogen-gas cooled furnace (Sigma 10M10J or Suns Electronic Systems EC1X). Other measurements were done at room temperature (290-298 K). A comparison of the effective piezoelectric coefficients was made by the sum over the resonance profiles of all resonance peaks (using IGOR PRO, WAVEMETRICS INC.). For samples with high peak intensities, RUS spectra of all compounds and RPS spectra in ferroelectrics, single crystal quartz, and in the paraelectric phases of BaTiO$_3$ and BZT20, the area of the spectrum was evaluated via integration, and the background was subtracted. In other materials, the total RPS area was calculated by fitting an asymmetric Lorentzian function to each resonance peak. RUS spectra were then used to scale the RPS spectra to estimate the effective piezoelectric coefficients (see section III). 

\begin{figure}
\includegraphics[width=12cm]{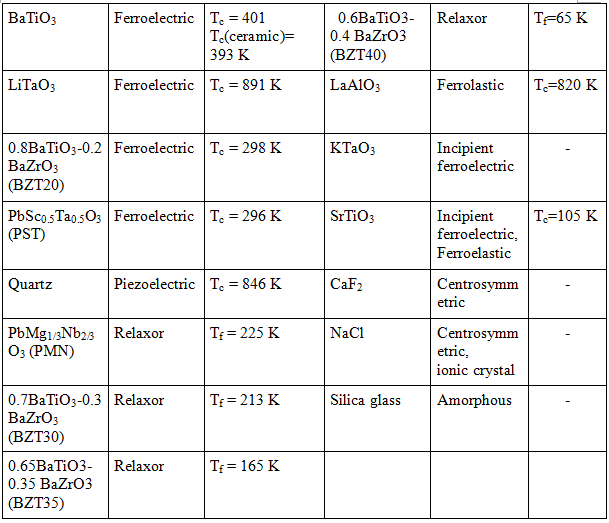}
\caption*{Table 1: Characteristics of compounds used in this work. Freezing temperatures, T\textsubscript{f}, associated with the freezing/slowing down of PNR’s in relaxors, are from Refs. \cite{maiti2008, maiti2006, shvartsman2009}.
}
\end{figure}

\section{ COUPLED RPS/RUS MEASUREMENTS AS A NEW METHOD TO MEASURE PIEZOELECTRICITY WITH HIGH RESOLUTION}

In this section, upon giving a brief discussion of the physical mechanisms behind RPS and RUS we introduce RPS/RUS measurements as a new method to measure effective piezoelectric coefficients of samples with high resolution. Then, we calibrate our results by using known ferroelectrics and paraelectric SrTiO$_3$. Other mechanisms that are allowed to contribute to the piezoelectric effect in RPS measurements were then evaluated on SrTiO$_3$.

\subsection{\textbf{\textit{Physical mechanisms behind RUS and RPS measurements:}}} The experimental arrangement for RUS and RPS is shown in Fig. 1(a). Both methods excite elastic standing waves (mechanical resonances) in the sample, whose frequencies are determined by the elastic moduli \cite{migliori1997, carpenter2015, aktas2013bto, salje2013sto}. In RUS, an AC voltage is applied across the piezoelectric emitter, which vibrates through the inverse piezoelectric effect \cite{migliori1997, carpenter2015}. These vibrations lead to strain oscillations in the sample, which is in contact with the emitter (Fig. 2). If the frequency of oscillations corresponds to one of the natural frequencies of the sample, the oscillations become elastic standing waves (i.e. mechanical resonances). The detection is then achieved by the piezoelectric detector via the direct piezoelectric effect. In RPS, the piezoelectric emitter is not used. Instead, the AC voltage is applied across the sample. The mechanical resonances of the sample are excited if the material is macroscopically piezoelectric \cite{aktas2013bto, salje2013sto, aktas2014kto, yokota2020, aufort2015}. Examples in the case of single crystal quartz, poled PZT-5H, and poled LiNbO3 are shown in Fig. 1. These resonances appear as peaks in both RPS and RUS spectra.

\subsection{\textbf{\textit{How to extract the piezoelectric response from the spectra:}}} In RPS, because the excitation of elastic resonances requires the sample to be piezoelectric, the area of resonance peaks gives a measure of the piezoelectric effect \cite{aktas2014kto, aktas2018nbt, aktas2018}. The area of peaks can be calculated simply by integration or fitting an asymmetric Lorentzian function to each resonance peak (see section II). 
\begin{figure}[htb]
		\includegraphics[width=6in]{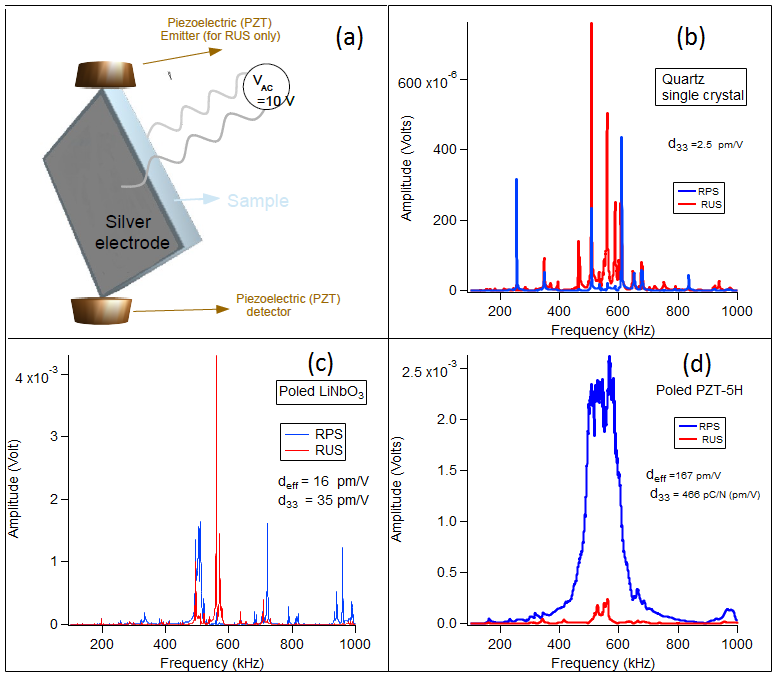}
		\caption{Elastic resonances of single crystal quartz, poled LiNbO$_3$, and poled PZT-5H measured piezoelectrically (by RPS) and mechanically (by RUS). (a) Schematic of RUS and RPS measurements. The sample is lightly held between two transducers. For RPS measurements, V\textsubscript{AC} = 1-20 V was applied across the sample and the resonances are detected with the piezoelectric detector. For RUS measurements, the AC voltage was applied on the emitter (instead of the sample) and the resonances are detected by using the same piezoelectric detector. RPS and RUS spectra of (b) quartz, (b) poled LiNbO\textsubscript{3} and (d) poled PZT-5H.}
\end{figure}

\subsection{\textbf{\textit{Examples of earlier works on the sensitivity of RPS:}}} RPS was specifically designed to detect miniscule piezoelectricity that cannot be measured by conventional direct piezoelectric measurements. The latter tends to have a resolution between 0.1 pm/V and 0.01 pm/V. The sensitivity of the RPS is because of the resonance condition of elastic standing waves \cite{aktas2013bto, salje2013sto, aktas2014kto, yokota2020, aktas2018, aufort2015}. This has previously been shown as observation of macroscopic piezoelectricity dominated by nano scale microstructures and defects in nominally centrosymmetric materials. These include piezoelectric twin walls in the ferroelastic phase of SrTiO3 \cite{salje2013sto}. The polar (consequently piezoelectric) nature of domain walls was later demonstrated by microscopic measurements \cite{frenkel2017}. Another example is coherent defect dipoles at cryogenic  temperatures in incipient ferroelectric KTaO3 \cite{aktas2014kto}. In this case, defect dipoles freeze in a coherent fashion, leading to an increase in piezoelectricity with decreasing temperature. In line with these measurements, complementary measurements showed spurious ferroelectricity in the same temperature range. Finally, in relaxor ferroelectric lead magnesium niobate PMN, piezoelectricity detected over a large temperature above room temperature was attributed to PNR's biased by chemically ordered regions \cite{aktas2018}. Recent observation of macroscopic polarization at room temperature confirms the piezoelectricity in PMN \cite{riemer2020}. The resolution of the technique was previously not explored. The lack of data on the (unexpected) piezoelectric coefficients of these centrosymmetric materials in the literature indicates that it has a higher resolution than 0.01 pm/V.  

\subsection{\textbf{\textit{Measurements of strain and effective piezoelectric coefficients:}}} To have access to piezoelectric coefficients that are beyond the resolution of direct piezoelectric measurements, we develop a method based on RPS and RUS. This is done as follows: 

\textbf{\textit{D1. Quartz as standard material and its piezoelectric properties:}}  Piezoelectric coefficients of single crystal quartz are reported by Bechmann as d\textsubscript{14} = 0.727 pm/V and d\textsubscript{11} = 2.31 pm/V, as reported by Bechmann \cite{bechmann1958}. The d\textsubscript{11} coefficient of the sample used in this work was d\textsubscript{11} = 2.5 $\pm$ 0.1 pm/V and is in line with the latter value. Note that our value is a result of direct piezoelectric measurements by a Berlincourt (d\textsubscript{33}) meter and not by RPS and RUS measurements. Small differences between reported values of d\textsubscript{11} naturally occur from sample to sample \cite{bechmann1958, bottom1970}. 

\textbf{\textit{D2. Calculation of average strain induced in RUS:}} Strain induced in quartz by the application of an electric field (i.e. in an RPS experiment) can be calculated by known piezoelectric coefficients of quartz through \textit{e}\textsubscript{RPS}{quartz} = \textit{d}\textsubscript{av} \textit{E}, where \textit{d\textsubscript{av}} is the average piezoelectric coefficient and \textit{E} is the electric field given by \textit{V}/\textit{t} with \textit{V} being the applied AC voltage and \textit{t} the sample thickness. Here, as both longitudinal and shear coefficients (d\textsubscript{11} and d\textsubscript{14}) contribute to the generated strain, we use d\textsubscript{av} = 1/2 (d\textsubscript{11}+d\textsubscript{14}). If the area of the RPS spectrum  (RPS (peak) area) designates the strain generated by the application of the AC field and the area of the RUS spectrum corresponds to the strain in an RUS experiment (\textit{e}$_{RUS}$), i.e. RUS (peak) area $\equiv$ \textit{e}$_{RUS}$, comparing the two areas, one can calculate the RUS strain
\begin{equation}
e_{RUS} = \mathrm{RUS (peak) area}* e_{\mathrm{RPS}}(\mathrm{quartz})/\mathrm{RPS (peak) area}
\end{equation}

\textit{The error in strain:} The strain value depends on the piezoelectric (d\textsubscript{14} and d\textsubscript{11}) coefficients (i.e. piezoelectric anisotropy) of quartz as well as the quality and geometry of the contact between the sample and transducers. To estimate the standard deviation in RUS strain, a total of 26 RPS/RUS measurements were carried out. An example of resulting spectra is shown in Fig. (2b). For every pair of RPS and RUS measurement, the sample was remounted to take into account the variations in local contact between the sample and transducers. The result is \textit{e}\textsubscript{RUS} = 3.5 $\times 10^{-7}$ with a standard deviation of 35\%. The calculated RUS strain spans the value of strain ($3\times 10^{-7}$) associated with a shear resonance generated during RUS measurements reported by Ref. \cite{ogi2004}.

\textbf{\textit{D3. Measurements and calibration of piezoelectric response to calculate the effective piezoelectric coefficients d\textsubscript{eff} of samples:}} This is done with a reverse procedure. Knowing the RUS strain (\textit{e}$_{RUS}$) and comparing the areas of RPS and RUS spectra (total areas of RPS and RUS peaks) of the sample, one can calculate strain generated in RPS, which then gives the effective piezoelectric coefficient through the equation \textit{e}$_{RPS}$(sample) = d\textsubscript{eff} \textit{E}. 
 
\textit{The error of effective piezoelectric coefficients:} Due to the range of values of the RUS strain, which is a result of piezoelectric anisotropy and quality and local geometry of contact between the sample and transducers, the effective piezoelectric coefficients determined with the approach adopted here should only be correct within an order of magnitude. Below, we test this by measuring the piezoelectric coefficients of poled ferroelectrics and parelectric SrTiO$_3$.

\textit{Effective piezoelectic coefficients of poled ferroelectrics:} Using the approach laid out above, the effective piezoelectric coefficients of poled PZT-5H and LiNbO$_{3}$ measured by RPS/RUS measurements (Figs. 2(c) and (d)) are d\textsubscript{eff} = 15 pm/V and  167 pm/V. The longitudinal piezoelectric coefficients of these samples measured by a Berlincourt (d\textsubscript{33}) meter are d\textsubscript{longitudinal} = 35 pm/V for LiNbO\textsubscript{3} and d\textsubscript{33} = 466 pm/V for PZT-5H.

\begin{figure}[htb]
		\includegraphics[width=6in]{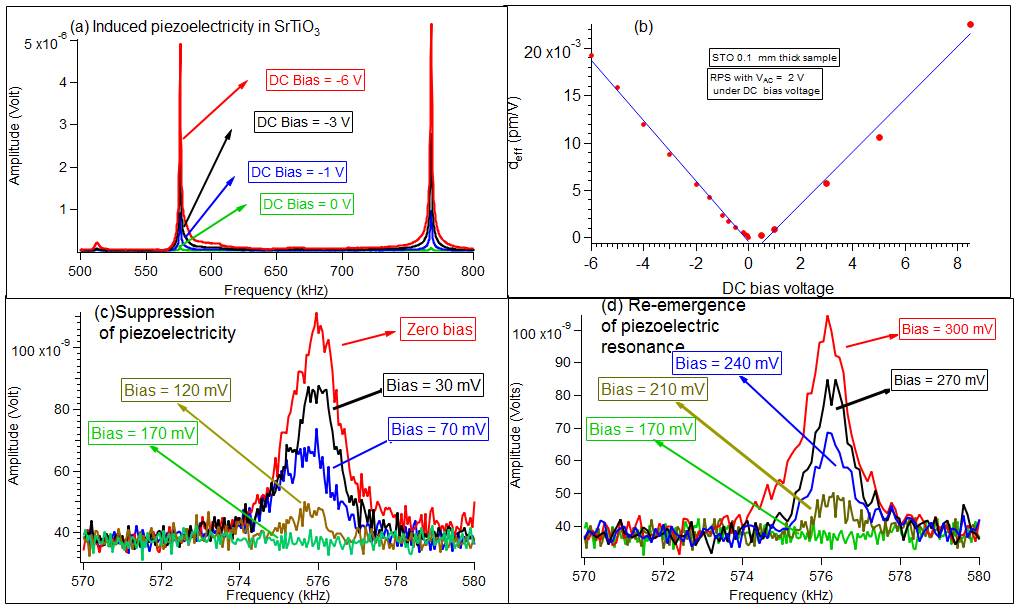}
		\caption{Calibration of RPS signal to determine ultra-low piezoelectric coefficients. (a) Induced piezoelectricity in SrTiO\textsubscript{3} by a DC voltage bias. (b) DC bias dependence of induced piezoelectricity. (c) Suppression of piezoelectric resonance by applying a positive bias (d) Re-emergence of the resonance by applying a bias greater than 170 mV.}
\end{figure}

\textit{The electrostrictive coefficient of paraelectric SrTiO$_3$:} Because a comparison between d\textsubscript{eff} measured by RPS/RUS measurements and direct d\textsubscript{33} measurements are not possible for values below 0.1 pm/V, i.e. the resolution of our d\textsubscript{33}, or below 0.01 pm/V for more sensitive d\textsubscript{33} meters, we use a different approach to calibrate the d\textsubscript{eff} values. Instead we determine the electrostrictive coefficient of SrTiO$_3$. This material is interesting not only on account of its archetypal status (the $``$drosophila of oxide physics$"$ , as K.A. Müller once called it), but also its structural, piezoelectric and electrostrictive properties are well documented. In addition, SrTiO\textsubscript{3} is an apparently perfect cubic perovskite above $T_c$ = 105 K, and any elastic softening above $T_c$ is associated with heterogeneities. It is therefore a perfect material for the study of nanopiezoelectricity in non-polar materials. 

SrTiO\textsubscript{3} has a very large dielectric constant ($ \varepsilon_r$=300 at room temperature), making it an easily polarizable material. This means that an electric field can be used to turn SrTiO\textsubscript{3} from a non-polar material into a polar one \cite{khan2016}, and then measure the $``$field- induced piezoelectricity$"$  as a function of electric bias. In an electrostrictive material such as SrTiO\textsubscript{3}, the strain is related to the polarization $P$ by \textit{e}=\textit{QP}\textsuperscript{2}, where $Q$ is the electrostrictive coefficient. The polarization in a perfectly non-polar crystal is the dielectrically induced polarization, $P=\varepsilon_0 \varepsilon_r E$, where \textit{E } is the electric field, $\epsilon_{0 }$is the permittivity of free space. In addition, however, there can also be a residual parasitic polarization coming from polar nanoregions, defect gradients and so forth; the total polarization is thus \textit{P=P\textsubscript{res}}+$\varepsilon_{0 } \varepsilon_{r}E$ with \textit{P\textsubscript{res}} being residual polarization. The effective $``$induced piezoelectric$"$  coefficient (\textit{d\textsubscript{eff}}\textsubscript{)} is, by definition, the derivative of the strain with respect to the electric field: 
\begin{equation}
d_{eff} =  \frac{\partial e} {\partial E} = \frac {\partial e \partial P} {\partial P \partial E} =  2 Q P \epsilon_0 \epsilon_r \\
= 2 Q \epsilon_0 \epsilon_r (P_{res} + \epsilon_0 \epsilon_r E)
\end{equation}
In a parallel-plate capacitor, the electric field is the voltage (\textit{V}) divided by thickness (\textit{t}). Hence,
\begin{equation}
d_{eff} =2 Q P \epsilon_0 \epsilon_r ((P_{res} +\epsilon_0 \epsilon_r \frac{V}{t})
\end{equation} 
The above equation means that we can linearly modify the piezoelectric coefficient of an SrTiO$_3$ single crystal by increasing the external bias voltage; the slope of the piezoelectric coefficient as a function of DC bias will be proportional to the electrostrictive coefficient and dielectric constant, while the intercept at the origin will be related to the residual macroscopic polarization of the sample. The result of this experiment, using an SrTiO\textsubscript{3} single crystal of 0.1mm thickness, is shown in Fig. 2. Notice that the induced piezoelectric coefficients are of the order of 10\textsuperscript{-3} pm/V (1 femtometer/volt) and would be hard to measure by standard techniques.

The slope of the RPS-measured piezoelectric coefficient of SrTiO\textsubscript{3} as a function of voltage is $ \sim $ 3.1 pm/V/m$^2$. From eq. (2) and assuming $\epsilon_r$=300 and \textit{t}=0.1mm, this slope implies an electrostrictive coefficient \textit{Q}=0.022 m\textsuperscript{4}/C\textsuperscript{2}, which is comparable to the longitudinal electrostrictive coefficient of SrTiO\textsubscript{3}, Q$_{33}$=0.0046m\textsuperscript{4}/C/m$^2$ \cite{khan2016}. This showcases the sensitivity of the RPS technique and its reliability, as well as offering an alternative way of calibration. Last, but not least, the measured piezoelectric coefficient at zero bias ($ \sim $ 2 x10$^{-4}$ pm/V) implies, via Eq. 2, a residual macroscopic polarization of the order of 1.7C/m$^2$. This is about 10$^5$ times smaller than the polarization of typical ferroelectrics such as BaTiO$_3$; conversely, if we assume that the detected residual polarization comes from coherently oriented polar clusters with a local polarization of the same order of magnitude as that of standard perovskite ferroelectrics, it means that the volume concentration of polar clusters in SrTiO$_3$ at room T is of the order of 10$^{-5}$ (10 parts per million). It would be difficult to observe such a small volume fraction of static polar clusters by direct means.

The limits of the RPS technique can be further tested if one considers that the forbidden piezoelectric coefficient is primarily a result of defect-gradient induced polarization \cite{biancoli2015, hashemi2016}. Therefore, by the application of a DC bias in the positive direction, we show that piezoelectric resonance can be suppressed and then emerge again. This means  switching the direction of residual polarization by a DC bias, further illustrating the accuracy and ultra-high sensitivity of RPS. The piezoelectric coefficient for small DC biases (120 mV and 210 mV) is calculated to be just below 10\textsuperscript{-5} pm/V. This sets the detection limit of RPS as 10\textsuperscript{-5 }pm/V, or 10 atto-meter per volt. 

\textit{Resolution and accuracy of the RPS method for piezoelectric measurements:} Our paper demonstrates that the RPS technique can detect piezoelectric signals down to 10$^{-5}$ pm/V. Though the resolution of our technique is unprecedentedly high, the margin of error is within an order of magnitude of the actual piezoelectric coefficient due to piezoelectric anisotropy and geometry and quality of local contact between the sample and transducers. This is well reflected in the extracted values of d\textsubscript{eff} for poled ferroelectrics and electrostrictive coefficient of SrTiO\textsubscript{3}, which differ from actual values by a factor of 3-5. However, this is sufficient for many scientific purposes, including the proof of existence of piezoelectricity in compounds that are nominally classified as non-piezoelectric.

\subsection{\textit{\textbf{Possible contributions of flexoelectric effect and surface piezoelectricity to RPS signals:}}} Here, both flexoelectricity \cite{abdollahi2019} and, theoretically, surface piezoelectricity \cite{hong2011, tagantsev2012,stengel2013, gattinoni2020} can generate strain that can potentially be picked in our measurements. Both are very weak effects in comparison to the piezoelectric effect on bulk crystals and ceramics \cite{bursian1968, zubko2013, tagantsev2012}. Nevertheless, we checked whether RPS can pick up these signals by performing size dependent measurements on the reference material SrTiO$_3$. The choice of SrTiO$_3$ as a reference material stems from the fact, in addition to being an apparently perfect paraelectric, its flexelectric properties have been extensively investigated \cite{zubko2007}. In fact, it is the first material for which the flexoelectric tensor was characterized \cite{zubko2007}. Unfortunately, for other materials in this study, the information is much less complete. In any case, it would be desirable to do similar studies in other materials, although it is out of the scope of this paper. Below we explain other possible contributions to the RPS signal that we have critically examined in the case of SrTiO$_3$:

\textit{Flexoelectric effect}: Bending experiments show that piezoelectricity can imitate flexoelectricity and vice-versa. Although in RPS measurements we do not bend the sample, we apply a homogeneous electric field that can induce bending via inverse flexoelectricity, and this may be mistaken as a piezoelectric signal \cite{abdollahi2019}. Therefore, we look at the size dependence of the RPS signal, comparing 0.1 mm and 1 mm thick samples (see supplementary material and Fig. 1). We see that both have piezoelectric coefficients on the same order of magnitude, which indicates that the origin is truly piezoelectric and not flexoelectric; if the origin of these piezoelectric coefficients had been inverse flexoelectricity, the thinner one would have displayed an orders-of-magnitude larger signal.
                          
\textit{Surface piezoelectricity}: Surface piezoelectricity is theoretically predicted in all materials \cite{hong2011, tagantsev2012,stengel2013, gattinoni2020}. By symmetry, surface piezoelectricity can only manifest itself in experiments where the equivalence between opposite surfaces is broken, i.e. when there is an inhomogeneous deformation such as bending so that one surface is under tension and the other under compression. Conversely, electric fields do break the equivalence between surfaces, because a field parallel to the piezoelectricity of one surface is antiparallel to the piezoelectricity of the opposite surface. This should cause a material to bend. Surface piezoelectricity is therefore indistinguishable from inverse flexoelectricity (this has in fact been amply discussed in the literature), and the same size-dependent experiment that allowed us to rule out a significant role of inverse flexoelectricity also excludes a role of surface piezoelectricity. In relation to this, it has been shown that surface effects are not the dominant mechanism for the forbidden polarization detected in paraelectric phases of ferroelectrics \cite{biancoli2015}. Nevertheless, surface piezoelectricity may make some (minor) contribution to the signal detected in piezoelectric measurements. These effects are likely enhanced in the case of rough surfaces, as shown by simulations \cite{lu2019}. 

\section{PIEZOELECTRICITY IN NOMINALLY CENTROSYMMETRIC PHASES OF COMPOUNDS}
\subsection{RPS and RUS spectra of nominally centrosymmetric and bulk-centrosymmetric materials}

Examples of RPS and RUS spectra for unpoled ferroelectrics and centrosymmetric materials are shown in Fig. 4 (see Fig. S1-S3 for the spectra of other compounds and samples). Similar to piezoelectric quartz, poled LiNbO\textsubscript{3},\begin{figure}[t!]
		\includegraphics[width=5in,height=6.74in]{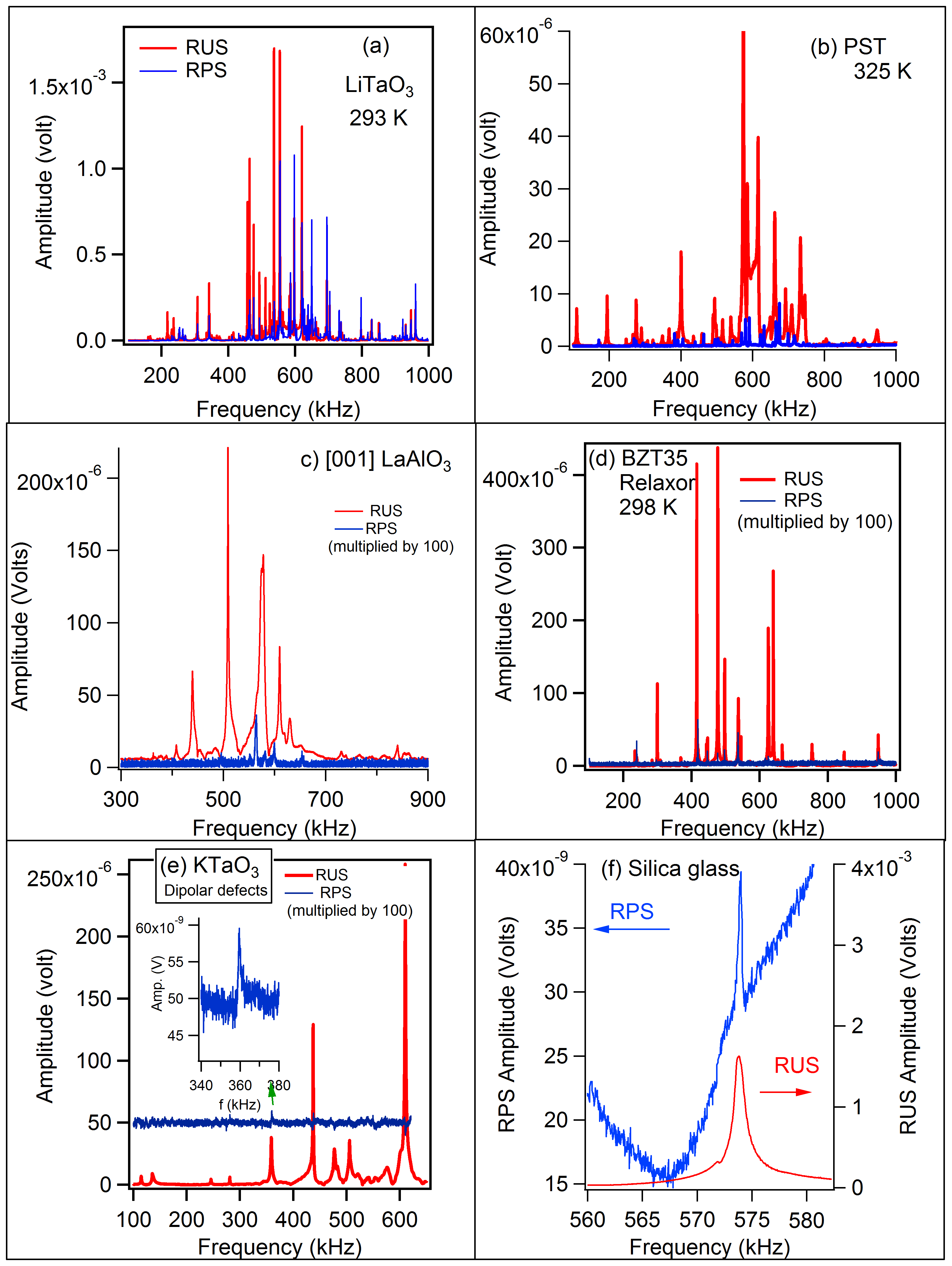}
		\caption{Piezoelectricity in globally centrosymmetric materials and unpoled ferroelectrics which are bulk-centrosymmetric due to spatial averaging of domains and grains (also see supplementary material). RUS spectra of all compounds show strong resonances while those detected in RPS range from 10\textsuperscript{-3} V to $ \sim $ 10\textsuperscript{-8} V, demonstrating broken centrosymmetry.}
\end{figure} and poled PZT-5H, all samples show elastic resonances in RPS spectra. Due to the disallowed nature of piezoelectricity in our unpoled ferroelectrics and nominally centrosymmetric compounds, the piezoelectric coefficients must be small when self-poling effects or intentional gradients (e.g. generated by heterogeneous chemical reduction) are absent \cite{he2017, zhou2015}. Biancoli et al. \cite{biancoli2015} measured the d\textsubscript{33} coefficient of a BaTiO\textsubscript{3} ceramic near the ferroelectric transition temperature and found that the coefficients were in the order of 0.1-0.3 pC/N (or pm/V). In our RPS measurements, BaTiO\textsubscript{3} as well as other ferroelectrics (Fig. 1 and Supplementary material), show strong resonance peaks that are only slightly below RUS peak amplitudes. In other samples, while all RUS spectra contain strong resonances with amplitudes of 10\textsuperscript{-3}-10\textsuperscript{-5 }V, RPS peaks are much weaker (Fig. 1 and supplementary material). This defines our range of values for the unexpected piezoelectric effect.

\subsection{\textit{\textbf{The magnitude of the observed piezoelectricity:}}} Using the approach described in section III, we calculated the effective piezoelectric coefficients of other samples (Fig. 1). 

In Fig. 1, d\textsubscript{33} of poled PZT-5H ceramic and LiNbO\textsubscript{3}, piezoelectric coefficients of quartz, are also shown for the purpose of depicting the range of piezoelectric coefficients that materials can have. The overall variation of piezoelectric coefficients spans an extremely broad range of values, varying by seven orders of magnitude from 466 pm/V to 10\textsuperscript{-5} pm/V (or 10 attometer/volt). These results somewhat reflects the range of deviation from inversion symmetry. 

Inversion symmetry is absent in poled ferroelectrics and quartz. Even unpoled ferroelectrics, although bulk-centrosymmetric, can give measurable piezoelectric coefficients\cite{biancoli2015} with conventional measurements but not always \cite{garten2015}, which also depends on the experimental resolution. Surprisingly, paraelectric phases of ferroelectrics lead to similar values that are only slightly lower (Fig. 1). This is consistent with the temperature dependence of d\textsubscript{33} for an unpoled BaTiO\textsubscript{3} ceramic reported by Ref. 5 which showed variations by a factor of 6 near T\textsubscript{c}. Thus, small differences of d\textsubscript{eff} below and above T\textsubscript{c} in Fig. 3 do not reflect the typical de-poling behavior and self-poling effects can be neglected. Taking quartz as reference, piezoelectric coefficients of centrosymmetric materials and silica glass (amorphous) range from comparable values to $ \sim $ 10\textsuperscript{-3} $\%$  of those of quartz (d\textsubscript{33} = 2.31 pm/V and d\textsubscript{14} = 0.727 pm/V \cite{bechmann1958}).

\subsection{\textbf{\textit{The role of polar nanostructures:}}} One should take the values of piezoelectric coefficients with caution especially for materials with local polar entities. Because the coherence between these structures is reduced on heating, one would expect the piezoelectric coefficient to also decrease with increasing temperature. As shown in Fig. 4, this is the case in the paraelectric phase of ferroelectric BZT20. This asserts earlier observations on BaTiO\textsubscript{3} \cite{aktas2013bto}, PMN \cite{aktas2018}, KTaO\textsubscript{3} \cite{aktas2014kto},\ and SrTiO\textsubscript{3} \cite{salje2013sto}, which led to the same conclusion based on RPS measurements.

\begin{figure}[htb]
		\includegraphics[width=3.62in,height=2.9in]{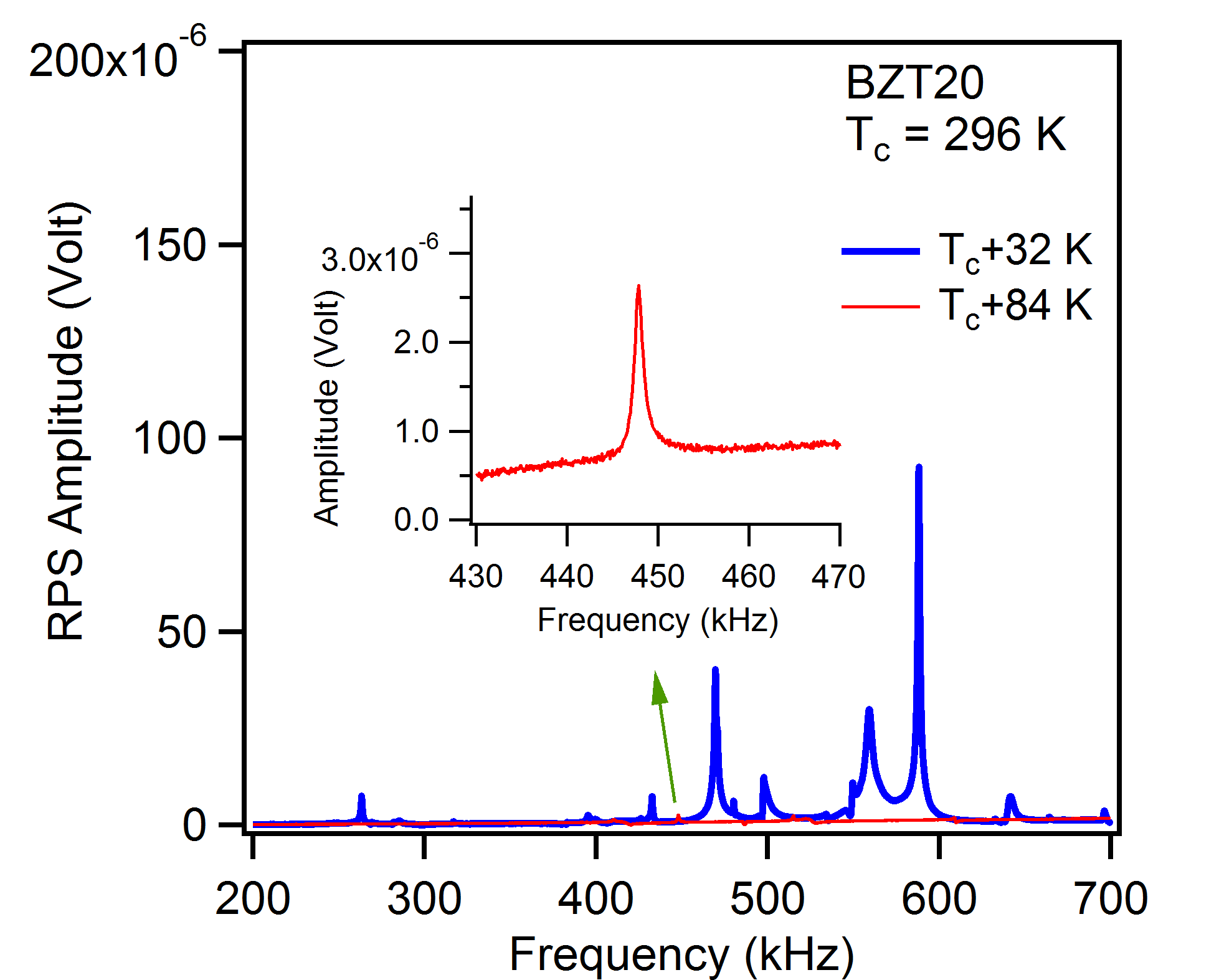}
		\caption{Comparison of RPS spectra of ferroelectric BZT20 collected 32 K and 84 K above the ferroelectric Curie temperature T\textsubscript{c} = 296 K.}
\end{figure}

Some of the compounds investigated in this work contain polar entities which have been visualized or inferred. In this context, we also include ferroelectric precursors, which occur in the paraelectric phase and exhibit locally the structural features which define the ferroelectric phase.\textsuperscript{11,27,32,33,74 }Polar entities in BaTiO\textsubscript{3}, its solid solutions and PST were discussed in detail \cite{bussmannholder2009, hashemi2016, garten2015, wieczorek2006,dulkin2010, bussmannholder2013, roleder2012, bussmannholder2014, ganlinyu2019,mihailova2008, aktas2013pst}. These have recently been visualized in BaTiO\textsubscript{3} \cite{benjan2020, tsuda2016} and its solid solution BaTiO\textsubscript{3}-SrTiO\textsubscript{3} \cite{benjan2020}. In the paraelectric phase of PST, striations observed in diffraction measurements \cite{babakishi1990, babakishi2010} were attributed to tweed structure via Landau modelling \cite{aktas2013pst} and scaling with entropy and polarization \cite{ganlinyu2019}. In BZT relaxors, PNR’s are estimated to persist at least up to 440 K \cite{shvartsman2009, maiti2008, maiti2006} while for PMN the hierarchical nano domain structure observed at room temperature potentially exists at least up to $\sim$600 K \cite{ eremenko2019, burns1973}. The polarity of twin walls in LaAlO\textsubscript{3 }and resulting macroscopic piezoelectricity was unambiguously demonstrated \cite{yokota2020}. Whereas materials with known polar nanostructures presented in this work have systematically higher piezoelectric coefficients than those with no known such structures at room temperature, it remains to be determined if such nanostructures are \textit{the cause} of symmetry breaking in all materials or an independent feature in some cases. 

\subsection{\textbf{\textit{Piezoelectricity in centrosymmetric materials which may contain external defects:}}}  Among other compounds investigated here, silica glass, NaCl, and CaF\textsubscript{2 }have no known locally polar or piezoelectric nano clusters but contain defects \cite{rabouw2016, silin1985, skuja2005, burow2009}. In KTaO\textsubscript{3 }and SrTiO\textsubscript{3} polar nanostructures only exists at cryogenic temperatures as polar clusters and polar domain walls, respectively \cite{salje2013sto, frenkel2017, uwe1986, aktas2014kto} No experimental evidence of local polarity for either sample has been reported at room temperature. Here, our results show that even cubic materials with no known polar or piezoelectric nanostructures are also piezoelectric. 

\section{CURRENT UNDERSTANDING OF SPONTANEOUS ATOMIC SCALE SYMMETRY BREAKING IN PARA PHASES}

The simplest, albeit naïve view of a para electric phase assumes that it has a net global zero dipole because each site has a zero dipole. This “non-electric model” of paraelectricity has been often used in electronic structure calculations as it allows one to use the smallest, highest symmetry crystallographic unit cell. Analogous approximations were common for describing paramagnets as a phase where each site has zero moment. Because it is artificial to expect a transition from finite dipoles in the low temperature ferroelectric phase to zero dipole in the paraelectric phase, a better approximation has been the displacive model of paraelectric phases \cite{merz1949} that still uses the minimal unit cell but allows for finite polar displacements in a single double well picture. However, because of the restriction to a minimal unit cell in the displacive model, all dipoles in the paraelectric phase must be aligned in tandem, leading to a long range ordered model for paraelectricity. In this view atoms are oscillating thermally in a single potential well around their average Wyckoff positions. Therefore, one can always apply time average onto the observable properties. Reciprocal-space calculation of phonons is based on such a symmetry unbroken monomorphous structure for the para phase following the softening of the para phonons as the transition to the lower temperature low symmetry phase is approached. In the alternative order-disorder model for the transitions \cite{bersuker1966} one allows in principle a larger than minimal unit cell for describing the para phase so that disordered local polar displacements are possible in the paraelectric phase, i.e. there is a potential well with a few minima (4 or 8) but the occupation numbers are variable allowing net polarity. Phase transition occurs when occupational symmetry is broken. However, the net dipole is zero.  

These concepts have been recently challenged \cite{hlinka2008}. Indeed, there are recent theoretical reasons \cite{zhao2020, zhao2021, wang2021, varignon2019, zhao2021bto} to believe that intrinsic mechanisms might be responsible for the formation of atomic-scale symmetry breaking in para phases, manifesting short-range-ordered (SRO) local motifs (“polymorphous network”) \cite{zhao2020}. Such symmetry breaking were argued theoretically to lower the internal energy U$_0$ even before thermal agitation sets in, leading to the formation of a \textit{distribution} of local motifs in all para phases. This includes para\textit{electric} phases (where the pertinent degree of freedom is the local dipole), or in para\textit{magnetic} phases (where the pertinent local degree of freedom is the local magnetic moment), or in para\textit{elastic} phase (where the pertinent degree of freedom can be a geometric-steric effect such as octahedral rotation). Significantly, even \textit{static minimization} of the internal density functional energy of supercells constrained to maintain the global symmetry already showed a significant stabilization in forming the polymorphous network that lets different local motifs coexist in one, large supercell. Naturally, as temperature sets in (modeled via DFT molecular dynamics (MD) \cite{zhao2021, zhao2021bto}), additions displacements take place. Significantly, both the minimization of the internal energy U (in DFT) and that of the free energy U-TS (via DFT-MD) lead to symmetry breaking including the removal of inversion symmetry. Thus, although  the \textit{average} <S> over local motifs $\{$$\mathrm{S}$\textsubscript{i}$\}$ has high symmetry (say, being centrosymetric), this does not imply that all measured physical properties P would equal the property P(<S>) of the average structure (e.g. null piezoelectricity). \textit{This then allows the observation of physical effects that reflect non-centrosymmetric symmetry, including piezoelectricity}. Recent first-principles calculations \cite{zhao2020, zhao2021, wang2021, varignon2019} considered a supercell of numerous Pm-3m unit cells, preserving the global cubic shape symmetry, but then minimizes the cell- internal atomic forces in Density Functional Theory (DFT). One finds in such constrained minimization of the internal energy U$_0$ a symmetry breaking. Depending on the type of pertinent local degree of freedom, these local motifs can correspond to local octahedra rotations in cubic halide perovskites (CsPbI$_3$) \cite{zhao2020, zhao2021}, or to local magnetic moments in paramagnetic oxides (YTiO$_3$, YNiO$_3$) \cite{varignon2019}, or to local displacements creating local dipoles in a paraelectric (BaTiO$_3$) \cite{zhao2020}. All such symmetry breakings represent short range order, SRO, and lead to macroscopic effects on the band structure, effective masses, optical properties \cite{zhao2020, zhao2021, wang2021, varignon2019}.  

\section{CONCLUSIONS }

Conclusions and potential future work can be summarized as follows:

\textbf{\textit{Macroscopic inversion symmetry breaking of cubic and bulk centrosymmetric materials:}}This was demonstrated to be ubiquitous. We demonstrated that unpoled ferroelectric crystals and ceramics can allow significant piezoelectricity.

\textbf{\textit{Internal vs external macroscopic symmetry breaking}:} We intentionally did not distinguish between internal \cite{salje2016flexo, zhao2020, zhao2021, wang2021, varignon2019, schiaffino2017} and external symmetry breaking \cite{biancoli2015, hashemi2016} on the detected piezoelectricity. Although, there is no experimental evidence of polar nanostructures in NaCl, silica glass, CaF\textsubscript{2}, KTaO\textsubscript{3}, and SrTiO\textsubscript{3} at room temperature, recent observations of macroscopic symmetry breaking in cubic materials due to internal effects\textsuperscript{75-78} suggest that bulk symmetry breaking could occur even in the absence of defects. For example, birefringence observed in CaF\textsubscript{2} has been proposed to stem from intrinsic effects \cite{burnett2001}. Future experimental and theoretical work would then include the quantification of intrinsic and extrinsic contributions to the symmetry breaking and piezoelectricity in nominally cubic materials. A possible approach to help distinguish the contribution of defects in symmetry breaking would require growing high-quality cubic crystals, such as NaCl, with known defect concentrations (some 10-1000 ppm). For examination of the local structure, transmission electron microscopy\cite{benjan2020, aert2012} are dark field x-ray microscopy\cite{simons2018} suitable methods. As macroscopic tools for investigation of inversion symmetry breaking, RPS, as shown in this work and earlier works \cite{salje2013sto, aktas2014kto, aktas2018, yokota2020}, and second harmonic generation \cite{fiebig2005}, as shown by Refs. \cite{fox1990, pugachev2012} could be carried out as a function of temperature on the same samples. 

\textbf{\textit{Applications:}} Possibility of new electromechanical devices based on paraelectrics. Piezoelectric coefficients in the paraelectric phases of ferroelectrics were found to be comparable to those of quartz and unpoled ferroelectrics. These results suggest that modifications of a collection of polar nanostructures may lead to new electromechanical devices. Such an attempt proved to be promising, for example, in Ref. \cite{zhou2015}, where introduction of heterogeneous chemical reduction leads to d$_{33}$ = 321 pm/V in Na$_{0.5}$Bi$_{0.5}$TiO$_3$-BaTiO$_3$, which exceeds the piezoelectric coefficients of most ferroelectrics and is comparable to that of lead zirconium titanate (PZT) \cite{gao2017}. Moreover, because the influence of polar nanostructures on the forbidden piezoelectric effect can be seen over hundreds of degrees above ferroelectric Curie temperatures and freezing temperatures associated with relaxors, high temperature piezoelectric applications, including energy harvesting, can be envisioned via the guided- growth/synthesis of materials via chemical, stress, thermal, and electrical gradients \cite{zhou2015, gerace2021, acosta2017}. 

\textbf{\textit{The scenario of intrinsic symmetry breaking short range order in para phases without defects and polar nanostructures:}} We know that \textit{positional} local symmetry breaking  such as displacements\ and octahedra rotations are seen by local structural probes  in nominally cubic perovskites (viz. pair distribution function PDF, Ref \cite{zhao2020}), whereas they often escape detection by volume-averaging techniques such as conventional XRD. Indeed, so are \textit{local} time reversal symmetry breaking predicted theoretically in paramagnets \cite{zhao2020, wang2021}, even when the global magnetism (vector sum of local moments) vanishes. Similarly, we see in the current work that, whereas the global dipole in a paraelectric could be small or vanishing, symmetry breaking calculations \cite{zhao2020} indicate that \textit{local} dipoles need not vanish. Remarkably, the appearance of such local motifs as energy lowering symmetry breaking features in paraelastics, paramagnets and paraelectrics can lead to macroscopic consequences, absent in the reference calculations lacking such local motifs. 

\section{SUPPLEMENTARY MATERIAL}
See the supplementary material for dimensions of all samples, orientations of single crystals, and the RPS and RUS spectra of samples not shown in the manuscript.

\section{ACKNOWLEDGEMENTS}
OA acknowledges the support of the Natural National Science Foundation of China (No. 51850410520). EKHS was funded by EPSRC (EP/P024904/1) and the EU’s Horizon 2020 programme under the Marie Skłodowska-Curie grant agreement No 861153. GC is funded by MINECO grant SEV-2017-0706 and the Generalitat de Catalunya grant 2017 SGR 579.Work of AZ at the University of Colorado at Boulder was supported by the USA National Science Foundation (NSF-DMR-CMMT Grant No. DMR-1724791) and the NDF DMREF program (Grant No. DMREF-1921949).

%



\end{document}